# An RFP dataset for Real, Fake, and Partially fake audio detection

Abdulazeez AlAli[1][0009-0009-9752-2481] and George Theodorakopoulos[1][0000-0003-2701-7809]

[1] Cardiff University, School of Computer Science and Informatics, Cardiff, United Kingdom
`aliaa8@cardiff.ac.uk, theodorakopoulosg@cardiff.ac.uk`

**Abstract.** Recent advances in deep learning have enabled the creation of natural-sounding synthesised speech. However, attackers have also utilised these technologies to conduct attacks such as phishing. Numerous public datasets have been created to facilitate the development of effective detection models. However, available datasets contain only entirely fake audio; therefore, detection models may miss attacks that replace a short section of the real audio with fake audio. In recognition of this problem, the current paper presents the RFP dataset, which comprises five distinct audio types: partial fake (PF), audio with noise, voice conversion (VC), text-to-speech (TTS), and real. The data are then used to evaluate several detection models, revealing that the available detection models incur a markedly higher equal error rate (EER) when detecting PF audio instead of entirely fake audio. The lowest EER recorded was 25.42%. Therefore, we believe that creators of detection models must seriously consider using datasets like RFP that include PF and other types of fake audio.

## 1   Introduction

Recent advances in machine learning and artificial intelligence have led to the development of more sophisticated forms of synthetic speech, including neural text-to-speech (TTS). These systems use deep learning algorithms to analyse and learn from large datasets of human speech, allowing them to produce speech that is more expressive and natural sounding. Recent research has demonstrated the ability to produce voice conversion (VC) and TTS by synthesizing voices for new speakers (not seen during training) called zero-shot [1]. Such developments have proven to be particularly beneficial when applied to various systems, including games, social media content, and the generation of voices to imitate the tone, texture, and pitch of someone's voice in films or TV shows. However, these developments also introduce opportunities for attackers to abuse synthetic speech. For example, they may use TTS technology to make malicious telephone calls that use the voice of a famous person for financial gain, replicate the voice of another to undertake phishing attacks or dupe a voice verification system by cloning somebody's voice. To deter such attempts, various detection models that enhance the efficiency of detection efforts by reducing the equal error rate (EER) have been developed [2, 3]. An EER is where the false rejection rate (FRR) and false acceptance rate (FAR) are equal [4].

An inspection of various synthetic and fake audio datasets reveals that each dataset contains a gap. Fake-or-Real dataset [5] includes a wide range of TTS synthetic speech. However, they are restricted to two data types: real and TTS. If someone needs to create a new detection model that can handle different types of fake audio, they should consider a separate dataset that contains VC, audio with noise and partial-fake (PF). PF is the process of replacing a segment within audio with a faked segment while the rest remains real. The PartialSpoof dataset [6] is based on the ASVspoof 2019 dataset. Although the PartialSpoof dataset includes PF utterances, it relies upon outdated TTS and VC generation techniques. In addition, there is no pair of real and fake utterances belonging to the same speaker, which is one of the biggest challenges of detecting fake audio. Another dataset called Half-Truth [7] delivers PF audio but does not include VC or audio with noise. Therefore, the partial fake is simply a combination of only real and TTS audio. Moreover, the dataset was created for the Mandarin language, similar to the ADD2022 [8] dataset. The ASVspoof challenge [9] produces multiple variants of the speech dataset. However, none of these versions contain PF audio, and the different sources of noise in the files were not within the scope of their research because the files contain only a low level of background noise. In [10], the effect of additive noise on spoofing detection performance was evaluated. With increased noise intensity, system performance was observed to decrease significantly. None of those datasets includes English PF audio within the completely fake or real audio dataset.

The RFP[1] dataset consists Real, Fake, and Partially fake audio. The dataset is intended for use to train new fake audio detection models as well as to assess the available detection models. The motivation for creating an RFP is to fill the gaps in existing datasets by generating a combination of PF audio. Inspired by [11] of having synthetic voices with equivalent real voices to be compared with. the RFP dataset contains real, TTS, and VC audio that we combine to generate PF audio. This results in the RFP dataset having two utterances from the same speaker reading the same text, one real and the other fake. The intention is that this dataset will be highly beneficial to researchers in several areas, including fake audio detection, replay attack, automatic speaker verification system (ASV), and training a voice for TTS.

The RFP dataset consists of five types of audio, one real and four fake, as follows:
• For real audio collections: the main source for real audio was obtained from YouTube-8m, which is an open-source project conducted by Google [12]. In addition, the [13] dataset was utilised to access a variety of different accents.
• For TTS generation: TTS and real audio were the initial PF audio combination. Although numerous TTS datasets are freely available, the decision was taken to produce new TTS files using a systematic method that addresses both genders, various accents, the latest TTS generation techniques and unique sentences for every individual utterance.
• For VC generation: The conversion process was undertaken using the latest open-source tools for voice conversion.
• Noise in audio: Having produced the VC and TTS conversion files, a selection of typical noises was added to the VC and real TTS audio files.
• For PF generation: Finally, every combination of real and audio with noise, real and VC, real and TTS, and PF audio was produced.
RFP includes 127,862 utterances from 354 speakers with distinct voices, of various ages, and from different regions.

---

[1] https://zenodo.org/records/10202142

Tests were then conducted by using a selection of synthetic speech detection models in an attempt to clarify how data generated by means of RFP affects the detection models' EERs. In doing so, it was confirmed that the detection models' EER was greater as a result of VC and PF. Meanwhile, the EER increased in only certain cases when applying TTS and audio with noise.

The remainder of this paper is organised as follows: Section II describes the creation of the RFP dataset and how the files were collected or generated, with an explanation of each tool used to create the dataset. Section III describes the two versions of the RFP dataset and the differences and the use of each version. Section IV provides an evaluation of the dataset using various detection models. Section V presents the experiment's results. Finally, Section VI covers the conclusion and future developments.

## 2  The RFP Dataset

There are five different types of audio in the RFP dataset: real, TTS, VC, audio with noise, and PF. Providing such varied audio utterances will help those developing fake audio detection models.

### 2.1  Real Audio Collection

A total of four sources were selected from a large number of freely available audio datasets: a high-quality, crowdsourced UK and Ireland English dialect speech dataset [13], the DiPCo - Dinner Party Dataset [14], the YouTube-8M Dataset [12] and the VCTK dataset [15]. These four datasets combine diverse settings, thereby helping to minimise unintended bias when evaluating existing or new detection models samples feature both genders, different accents, noisy and noise-free settings, a range of microphones, and both low- and high-quality audio.

Crowdsourced high-quality UK and Ireland English Dialect speech dataset [13] is an open-source dataset which includes 18,779 utterances spoken by 118 individuals. We chose this dataset because those who participated came from six different regions in the UK and Ireland and spoke different varieties of English, namely Irish, Midlands, Northern, Scottish, Southern and Welsh English.

The DiPCo - Dinner Party Dataset [14] is multiple four-person groups of volunteer Amazon employees engaged in natural English conversation at a dining table to create the corpus. The recording hardware was a single-channel close-talk microphone, and five far-field 7-microphone array devices were strategically situated in the recording chamber. Because the recording sessions can last up to 45 minutes and contain numerous silences and laughing, the Pydub Python library was used to remove them and only utterances comprising mostly of the talk were retained, resulting in a total of 2,022 utterances spoken by 20 speakers.

The YouTube-8M Dataset [12] provided by Google is an open-source subset of YouTube. This dataset was chosen because it comprises millions of videos and audio clips from a wide range of recording and environmental scenarios. Examples of audio taken from this dataset and included in the RFP dataset include a conference video with background noise, high-quality educational videos such as TED lectures, and street interviews. 95 males and 102 females were selected. A total of 9,717 utterances were collected, edited and prepared for machine learning usage.

Even though the participants in the VCTK dataset [15] have various accents, the motivation for using it was the content used for recording: rainbow passage and other text collected from multiple sources. Four males and three females with a total of 4,943 utterances were selected from this dataset. Table 1 presents statistics and a list of selected real audio datasets.

**Table 1.** Real Data Statistics

| Dataset | Speakers | | Utterances | | Total |
|---|---|---|---|---|---|
| | M | F | M | F | |
| UK and Ireland | 69 | 49 | 10627 | 7250 | 17877 |
| YouTube-8M | 81 | 113 | 3416 | 6296 | 9712 |
| DiPCo | 10 | 7 | 398 | 128 | 526 |
| VCTK | 4 | 3 | 3156 | 1952 | 5108 |
| Total | 164 | 172 | 17597 | 15626 | 33223 |

### 2.2 Text-to-Speech Audio Generation

TTS has become crucial in a variety of applications and usages; however, it is vulnerable to attackers who can abuse the TTS to carry out attacks. Attackers use the most realistic TTS projects to create convincing audio, which has resulted in the selection of seven cutting-edge TTS cloud services and open-source applications. Before starting to generate TTS audio, it is necessary to have sufficient unique sentences that can be read by TTS tools. At first, 34,000 unique sentences from the British National Corpus (BNC) [16] were used. The strength of this corpus is that it contains 100 million words from a wide range of genres (such as spoken word, fiction, magazines, newspapers, and academic papers). The sentences were shortened, duplicates were removed, and brief sentences (those containing four or fewer words) were eliminated. Table 2 presents the statistical data and models used to create the text-to-speech (TTS) audio files.

Amazon Polly[2] is a text-to-speech conversion service that operates in the cloud. It generates natural-sounding speech by utilising powerful deep learning technology. There are two types of TTS available: neural and standard. Each of these features some male and female voices from various places with different accents. To begin, an Amazon AWS account was created to gain access to the Amazon Polly TTS service. A script was created, and the integration was completed. A text file containing sentences was transmitted to Amazon Polly and spoken utterances were obtained. Four speakers were chosen to create a total of 8,000 audio files, 2,000 of which were created by each. The four speakers had American, British, Australian, and Welsh accents.

Windows Azure[3] is similar to Amazon Polly and each stage was completed with an equal number of generated utterances but with different accents from multiple countries (Ireland, Canada, Singapore and New Zealand) with gender balance: two males and two females. In addition, custom neural and neural voices are the two TTS varieties offered by Windows Azure.

Google Cloud TTS[4] and gTTS[5] : There are three distinct varieties of Google Cloud TTS: neural, wavenet, and standard voices. Six speakers were chosen and a total of 12,000 audio files were generated by employing Amazon Polly and Windows Azure-like procedures. In addition, Google offers a second TTS service called gTTS, a Python library which is utilised by Google Translate. Unlike other cloud-based TTS services, registration is not required to use gTTS. However, it restricts the number of requests that can be transmitted per minute.

FastPitch [17] is a fully parallel open-sourced TTS system that predicts pitch contours during inference and produces speech that can be fine-tuned using these predictions. 2,000 utterances were generated using FastPitch.

The Coqui TTS[6] Toolkit is a set of tools for advanced TTS generation based on deep learning. It is based on the most recent research and was designed to achieve the optimal balance between training simplicity, training speed and training quality. Two deep learning models were chosen to produce 2,000 utterances overall: Glow-TTS [18] and Tacotron 2 [19].

---

[2] https://aws.amazon.com/polly/
[3] https://azure.microsoft.com/en-us/products/cognitive-services/text-to-speech/
[4] https://cloud.google.com/text-to-speech
[5] https://github.com/pndurette/gTTS
[6] https://github.com/coqui-ai/TTS

**Table 2.** TTS data statistics

| Model | Speakers | | Utterances | | Total |
|---|---|---|---|---|---|
| | M | F | M | F | |
| Amazon AWS Polly | 2 | 2 | 4000 | 4000 | 8000 |
| Windows Azure | 2 | 2 | 4000 | 4000 | 8000 |
| Google Cloud TTS | 3 | 3 | 6000 | 6000 | 12000 |
| Google gTTS | - | 1 | - | 2000 | 2000 |
| FastPitch | - | 1 | - | 2000 | 2000 |
| GlowTTS | - | 1 | - | 1000 | 1000 |
| Tacotron2 | - | 1 | - | 1000 | 1000 |
| Total | 7 | 11 | 14000 | 20000 | 34000 |

### 2.3 Voice Conversion (VC) Audio Generation

VC is the process of altering a voice signal from a source speaker to sound as though it was said by a target speaker while maintaining the original linguistic content.

VC employs various strategies, including one-to-one, one-to-many, and many-to-one. It can also be performed by individuals of various genders and languages. Because this form of service raises privacy concerns, the available services require personal verification. Consequently, four open-source VC projects were used to generate 18,000 audios using the approaches. The models and statistics used to generate VC audio files are listed in Table 3.

Soft-VC [20] converts any source speaker into a single target speaker. The content encoder extracts discrete or soft speech units from input audio, adhering to the preceding stages' standards. The acoustic model converts speech units into the desired spectrogram. The vocoder then transforms the spectrogram into an audio waveform. Five sets of conversions have been performed using Soft-VC, generating a total of 4,000 outputs from three distinct TTS and two real audio sources.

Free-VC [21] is an end-to-end framework for high-quality waveform reconstruction that is proposed along with strategies for extracting content information without text annotation. 6,000 VC utterances and conversions were generated using this tool based on one-to-one, many-to-one, any-to-one and male-to-female voices.

PPG-VC [22] is an any-to-many location-relative, sequence-to-sequence (seq2seq), non-parallel voice conversion strategy proposed in the current study and uses text supervision during training. This method uses a seq2seq synthesis module and a bottleneck feature extractor (BNE). It achieves superior vocal conversion performance in terms of naturalness and speaker similarity. Four conversions were performed to produce 4,000 utterances.

Diff-VC [23] is unlike many VC tools that require the target speaker to be part of the training dataset because it supports the most typical scenario; when both the source and target voices do not belong to the training dataset, a one-shot many-to-many voice conversion replicates the target voice from only one reference utterance. 4,000 utterances were generated using this tool.

A VC algorithm takes two audio files, source and target, as input and generates the converted voice as output. This output converted audio is similar to the input target audio in the sense that it says the same words as the source audio, but it sounds like the person who speaks in the target audio. This conversion approach will present a challenge on the detection side because the detection needs to evaluate files sharing the same content.

**Table 3.** VC data statistics

| Model | Source Speakers | | Target Speakers | | Utterances | | Total |
|---|---|---|---|---|---|---|---|
| | M | F | M | F | M | F | |
| Soft-VC | 11 | 14 | - | 1 | - | 4894 | 4894 |
| Free-VC | 9 | 12 | 4 | 2 | 4000 | 2000 | 6000 |
| PPG-VC | 8 | 7 | 3 | 1 | 3000 | 1000 | 4000 |
| Diff-VC | 18 | 21 | 3 | 1 | 3000 | 1000 | 4000 |
| Total | 46 | 54 | 10 | 5 | 10000 | 8894 | 18894 |

### 2.4 Audio With Noise Generation

Almost everywhere in the real world, various types of noise exist. For example, you may call a bank while you are at home and the washing machine is on or answer an important call in the office while someone is loudly typing on a keyboard. The attacker may use the same concept of background noise while impersonating a person's voice to carry out an attack. Therefore, noise cannot be disregarded by detecting synthetic speech tools. As such, 12 different noises (six internal and six external) were chosen and applied to Real, TTS, and VC audio files. All of the selected noises differed in A-weighted decibel (dBA) level, which measures the loudness of sounds as perceived by the human ear. The selected noises ranged from 50dBA (equal to the noise emitted by a washing machine) to 140dBA (equivalent to an aeroplane taking off). These different types of noises are important to analyse during the detection process to provide efficient and robust fake audio detection models. The noises sounds were collected from the Freesound[7] website.

---

[7] https://freesound.org/

In order to merge the noise with the required data, we created a Python script that used Librosa[8], a Python package for music and audio analysis. To make the audio more realistic and understandable after merging, the gain of the noise audio, which is measured in decibels (dB), was adjusted to an acceptable level to enable the person to understand the content of the audio. The amount of gain adjustment was based on the type of noise. Table 4 shows which audio types are merged with selected indoor and outdoor noises.

**Table 4.** Audio with noises data statistics, the total number of utterances is 12000.

| Noise | Indoor/Outdoor | M/F | Type of file |
|---|---|---|---|
| Airplane Taking off | Outdoor | M | TTS |
| Ambulance Siren | Outdoor | M | TTS |
| Car Pass | Outdoor | F | TTS |
| Crowd | Outdoor | F | VC |
| Rain | Outdoor | M | Real |
| Wind | Outdoor | F | Real |
| Baby crying | Indoor | F | TTS |
| Coffee grinder | Indoor | M/F | VC |
| Gathering | Indoor | M | VC |
| Keyboard | Indoor | F | TTS |
| Vacuum | Indoor | M | TTS |
| Washing machine | Indoor | M | TTS |

### 2.5 Partial Fake (PF) Audio Generation

Both fake and real audio files are required to produce PF audio. The PF audio for the RFP dataset involved the use of VC, TTS, audio with noise, and real audio files. PF audio is produced using both genders and different audio types. PF was generated based on four concatenation cases as set out below:

• Real-Fake-Real: A total of three parts, beginning with real, then fake, and final-ly, a real part.
• Fake-Real-Fake: A total of three parts, beginning with fake, then real, and final-ly, a fake part. Figure 1 presents TTS-Real-TTS, which is an example of Fake-Real-Fake.
• Fake-Real: A total of two parts, beginning with fake and finally real.
• Real-Fake: A total of two parts, beginning with real and finally fake.

This combination of PF is applied to consider the full range of possibilities for incorporating fake audio into real audio. Such an approach will help to determine how efficiently detection models operate.

This combination of PF is applied to consider the full range of possibilities for incorporating fake audio into real audio. Such an approach will help to determine how efficiently detection models operate. The statistics for various PF combinations are listed in Table 5.

---

[8] https://librosa.org/

**Table 5.** Partial fake data statistics

| Model | Speakers | PF Combination | Utterances |
|---|---|---|---|
| TTS | Female | Real-TTS | 1412 |
|  |  | TTS-Real | 1000 |
|  |  | Real-TTS-Real | 2820 |
|  |  | TTS-Real-TTS | 2500 |
|  | Male | Real-TTS | 1624 |
|  |  | TTS-Real | 2000 |
|  |  | Real-TTS-Real | 3059 |
|  |  | TTS-Real-TTS | 1584 |
| VC | Female | Real-VC | 988 |
|  |  | VC-Real | 1963 |
|  |  | Real-VC-Real | 2817 |
|  |  | VC-Real-VC | 1849 |
|  | Male | Real-VC | 1328 |
|  |  | VC-Real | 1807 |
|  |  | Real-VC-Real | 2888 |
|  |  | VC-Real-VC | 1666 |
| **Total** |  |  | 31305 |

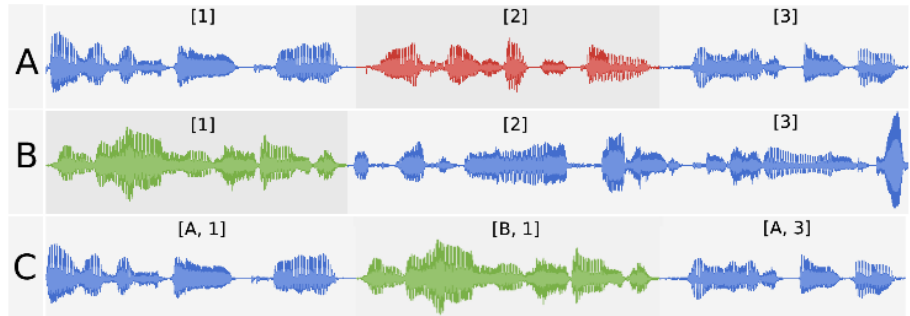

**Fig 1.** An example of partial fake generation in the case of TTS-Real-TTS, where A is TTS audio, B is real audio, the red portion of the wave on A[2] is replaced with the green portion of B[1], and the final partial fake audio is shown in C represent A[1], B[1] and A[3].

## 3 Dataset Composition

The RFP dataset is available in two distinct versions. The original is available in the form that it was generated or collected in without being processed. It can be used for a range of purposes, such as to train new TTS voices or to conduct replay attacks. Meanwhile, the normalised version is balanced, has undergone processing, and can be directly applied for machine learning.

### 3.1 RFP Original Version

The audio files in the original version are unmodified, and the file formats include WAV, MP3 and FLAC. The sampling rates of the files range from 16kHz to 48kHz, and any files generated by open-source software or scripts we developed use 16 kHz mono WAV files. The names of the folders and subfolders indicate the origin of each file produced, including the PF folder, which contains the file combinations used in each generation technique. In addition to the source and target speakers used for voice conversion, this version includes transcripts for most of the TTS and real voice files to assist the TTS voice training model. The whisper model [24] was used to perform the transcription. This version consists of a total of 127,862 utterances spoken by 354 speakers, of which 184 are male and 170 are female.

### 3.2 RFP Normalised Version

This version processes and prepares the data for use with machine learning. Everything stereophonic in the audio was converted to monophonic (mono) audio signals, which consist of a single channel. By contrast, stereo audio signals contain two channels that depict a 'left' and 'right' that correspond to what is heard on the speakers during playback. Second, because different sample rates exist for audio, all sample rates were converted to 16kHz. Third, all audio was converted to the preferred format for audio processing in machine learning, which is WAV. Finally, the gender distribution and total number of file categories were balanced. In practice, it is impractical for audio files to be the same length, despite the fact that some detection models necessitate it. Therefore, the duration of these dataset version files ranges from 3 to 20 seconds. This version contains 108,385 balanced utterances from 115 male and 115 female speakers.

This normalised version is divided into three sections, adopting the suggested division [25] provided. Two-split (training and testing) data is not an efficient practice because it leads to snooping, another type of overfitting. Snooping can lead to overfitting in several ways: Data Leakage, Optimizing for Test Set Performance, and Feature Engineering and Selection. One of the best ways to avoid overfitting due to snooping is to strictly separate your data into training, validation, and test sets. The third set (validation) helps to prevent snooping. Each set of the dataset comprises unique speakers or utterances. The division is as follows:
- Training: This set contains 76,796 utterances, which represents 70.9% of this RFP version. The purpose of training is to optimise the model's parameters so that the difference between the predicted and actual output is as small as possible.

- Validation: This set contains 21,611 utterances, or 19.9% of the dataset. Validation can assist in detecting overfitting and adjusting the model's hyperparameters to enhance its generalisation performance.
- Testing: This set contains 9,978 utterances, or 9.2% of the dataset. The test set is utilised to evaluate the generalisation performance of the model as well as to estimate its performance on new, unseen data.

## 4    Experiments

The normalised test set was employed to conduct several evaluation tests and examine how effective the fake audio is to existing fake audio detection models. Particular attention was paid to the effect that partially fake audio has on the detection models' EER. Motivated by the ASVspoof 2021 challenge [9], the four provided baselines were utilised to compute the EER. In addition, the three tools with the lowest EER provided by teams participating in ASVspoof challenges were utilised. These detection models offered a comprehensive picture of the robustness and efficacy of these tools in detecting unseen and partially fake audio. All tools were converted to Jupyter Notebook versions, and the experiments were conducted using the Google Colab platform. In addition, all of the published results for the logical access (LA) dataset to the baselines and the tools supplied by the three teams were regenerated to verify the detection's validity and compared with our results using the RFP dataset.

### 4.1    ASVspoof 2021 baseline

The Automatic Speaker Verification Spoofing and Countermeasures (ASVspoof) Challenge is a biennial international competition concerned with spoofing attacks on ASV systems. Spoofing attacks aim to fool an ASV system into believing that a recorded voice sample is the voice of a legitimate speaker when it is actually a fake or 'spoofed' voice created by an attacker. The ASVspoof Challenge provides a standardised framework for comparing the effectiveness of various ASV deception countermeasures. The four baselines for the ASVspoof 2021 release are as follows.

- CQCC-GMM: Constant Q cepstral coefficients (CQCC) [26] feature extraction with Gaussian mixture model (GMM) classifier.
- LFCC-GMM: Linear Frequency Cepstral Coefficients (LFCC) [26] feature extraction with GMM classifier.
- LFCC-LCNN: LFCC feature extraction with light convolutional neural network (LCNN) [27] classifier deep neural network (DNN).
- RawNet2: End-to-End DNN classifier [28].

### 4.2 TSSDNet

This is an end-to-end synthetic speech detection system which employs two lightweight neural network models called time-domain synthetic speech detection net (TSS-DNet) with the classic ResNet and Inception Net style structures (Res-TSSDNet and Inc-TSSDNet) [29]. The two models were included in the experiment.

### 4.3 RawGAT-ST

This is a spectro-temporal graph of attention networks end-to-end for speaker verification. Anti-Spoofing and Speech Deepfake Detection illustrates the efficiency of an end-to-end spectro-temporal graph attention network (GAT) for anti-spoofing and speech deepfake detection by learning the link between cues spanning distinct sub-bands and temporal intervals [30].

### 4.4 AASIST

In this project, they propose a novel heterogeneous stacking graph attention layer that uses a heterogeneous attention mechanism and a stack node to model artefacts across diverse temporal and spectral domains. The two proposed architectures were used to evaluate the datasets AASIST and AASIST-L [31].

## 5 Experiments Result

Table 6 presents the EER results of nine different detection models using the RFP normalised test set for distinguishing between real and TTS audio, the LFCC-GMM detection model exhibits an EER of 0%, which is superior to all other models. For the audio types of real and VC, real and noise, and real, VC and TTS, as well as for all audio types with and without PF, LFCC-LCNN detection methods achieved the best EER results.

Finally, PF is the audio type that is the hardest to detect, in the following sense: The best EER for PF is 25.42% (achieved by RawGAT-ST), whereas the best EERs for all other audio types are lower than 25.42%, sometimes as low as 0%. Many of the detection models could detect certain types of fake audios but were deceived by others. For instance, LFCC-GMM recorded 0% for TTS, but surged to 71.23% for audio with noise. Meanwhile, RawNet2 recorded 7.9% for TTS and 55.62% for VC.

As such, it is apparent that more effective and reliable detection models capable of detecting the full range of fake audio types are required. In every test, CQCC-GMM was the worst-performing detection model.

**Table 6.** THE EVALUATION RESULTS FOR ASVSPOOF 2021 BASELINES USING THE LA DATASET AND FIVE DETECTION METHODS. RESULTS SHOWN IN TERMS OF POOLED EER [%].

| Detection Method | ASVSpoof 2021 dataset | RFP Dataset EER [%] | | | | | | |
|---|---|---|---|---|---|---|---|---|
| | | Real, TTS | Real, VC | Real, Noise | Real, PF | Real, VC, TTS | All except PF | All |
| LFCC-GMM | 15.62 | 0 | 49.46 | 71.23 | 59.49 | 29.17 | 46.37 | 55.49 |
| CQCC-GMM | 19.30 | 80.64 | 60.12 | 84.28 | 64.23 | 75.26 | 77.74 | 71.23 |
| LFCC-LCNN | 9.26 | 5.94 | 23.79 | 11.10 | 34.71 | 16.45 | 15.13 | 22.74 |
| RawNet2 | 9.50 | 7.90 | 55.62 | 25.71 | 37.46 | 27.27 | 26.56 | 32.02 |
| Res-TSSDNet | 1.64 | 27.42 | 28.14 | 33.77 | 27.59 | 27.63 | 29.22 | 28.43 |
| Inc-TSSDNet | 4.03 | 25.02 | 26.91 | 25.12 | 25.97 | 25.58 | 25.46 | 25.71 |
| RawGAT-ST | 1.39 | 5.21 | 43.79 | 16.54 | 25.42 | 21.90 | 19.24 | 28.97 |
| AASIST | 0.83 | 15.80 | 41.30 | 20.08 | 26.14 | 23.68 | 21.98 | 29.01 |
| AASIST-L | 0.93 | 18.45 | 49.96 | 23.00 | 30.33 | 27.94 | 25.13 | 35.90 |

# 6 CONCLUSIONS AND FUTURE DEVELOPMENTS

RFP dataset is one of only a few freely available English datasets that includes PF audio for use by those interested in detecting fake speech. The state-of-the-art generation techniques were used to produce the VC and TTS files, thereby ensuring they sound as natural as possible. The dataset contains more than 127,000 utterances. There are two versions of the dataset: The original version contains the actual files as collected or generated and can be used in various applications. The second version is processed, prepared, and divided into training, validation, and testing sets, which can be used in any machine learning detection model. Using the new dataset, we evaluated nine detection models, and the results demonstrate the critical need for an enhanced or new detection model capable of handling all sorts of faked and PF audio. The detection model should also be sufficiently robust to deal with unseen data.